\documentclass[runningheads]{llncs}
\usepackage[T1]{fontenc}
\usepackage{graphicx}

\usepackage{zref-clever} 
\zcsetup{cap,nameinlink,abbrev}

\usepackage[colorlinks=true,allcolors=purple]{hyperref}

\newcommand{\R}{\mathbb{R}}

\usepackage{float}
\usepackage{amsmath}
\usepackage{xcolor}
\usepackage{booktabs}
\usepackage{subcaption}
\usepackage{amssymb}
\usepackage{amsmath}
\usepackage{makecell}
\usepackage{listings}
\usepackage[shortlabels]{enumitem}
\usepackage{tikz}

\usepackage{wrapfig}

\usepackage[linesnumbered,ruled]{algorithm2e}
\SetKw{Real}{real}
\SetKw{Int}{int}
\SetKw{Either}{either}
\SetKw{Or}{or}

\usepackage{bbding}

\begin{document}
	\title{Fast Ramsey Quantifier Elimination in LIRA \\
		(with applications to liveness checking)}

	\titlerunning{Fast Ramsey Quantifier Elimination in LIRA}
	%
	\author{Kilian Lichtner\inst{1} \and
		Pascal Bergsträßer\inst{1} \and
		Moses Ganardi\inst{1} \and
		Anthony W. Lin\inst{1,2} \and
		Georg Zetzsche\inst{2}}
	\authorrunning{Lichtner et al.}
	%
	\institute{RPTU University Kaiserslautern-Landau
		\\\email{\{lichtner,pbergstr,moses.ganardi\}@rptu.de}
		\and
		Max Planck Institute for Software Systems (MPI-SWS)
		\\\email{\{awlin,georg\}@mpi-sws.org}
	}

	\maketitle              

	\begin{abstract}
Ramsey quantifiers have recently been proposed as a unified framework for
handling properties of interests in program verification involving
proofs in the form of infinite cliques, which are not expressible in 
    first-order logic. Among others, these include liveness verification and 
    monadic decomposability.
We present the tool REAL, which implements an efficient elimination of
Ramsey quantifiers in existential linear arithmetic theories over integers 
(LIA), reals (LRA), and the mixed case (LIRA). The tool supports a convenient
input format, which is an extension of SMT-LIB over the aforementioned
theories with Ramsey quantifiers. We also demonstrate a substantial speedup from
the original prototype. As an application, we provide an automatic translation
from FASTer (a tool for verifying reachability over infinite-state systems) output format to our 
extension of SMT-LIB and show how our tool extends FASTer to liveness checking.

\end{abstract}

	\section{Introduction}
	\label{sec:intro}
	
	Satisfiability Modulo Theories (SMT) (cf.~\cite{SMT}) is a highly successful 
automated 
reasoning framework that is based on first-order theorem proving. In particular,
building on the success of propositional SAT-solvers, SMT-solvers support 
numerous first-order theories including linear arithmetics, array 
theory, string theory, bitvector theory, and Equality logic with Uninterpreted
Functions (EUF), to name a few. The SMT framework is also suitable for reasoning
about programs with respect to safety properties, e.g., using Constraint Horn
Clauses \cite{GLPR12,BMR12}. Proving liveness (e.g. termination), however,
typically requires an extension of the SMT framework with reasoning about
well-foundedness, which is not first-order definable even with the help of
recursive predicates (equivalently, second-order quantifiers). 

In recent years, Ramsey quantifiers have been proposed
\cite{ramseyquantifierslineararithmetics,ramsey-aut} as a uniform framework for handling
properties of interests in program verification involving infinite
\emph{cliques}. These include, among others, liveness \cite{TL08,TL10,ramsey-aut} and 
monadic decomposability \cite{mondec}. In essence, we extend an SMT with formulas of the form
\begin{equation}\label{eq:ramsey}
    \psi(\vec z) := \exists^{\text{ram}} \vec x, \vec y \colon 
        \varphi(\vec x,\vec y,\vec z)
\end{equation}
with $|\vec x| = |\vec y|$ and with the semantics that it is true under the 
valuation $\vec z \mapsto \vec c$ iff the graph $\varphi(\vec x,\vec y,\vec c)$
has an infinite directed clique, i.e., there is
an infinite sequence $\vec a_1, \vec a_2,\dots$ of pairwise distinct vectors 
such that $\varphi(\vec a_i,\vec a_j,\vec c)$ holds for all $1 \le i < j$.
To see how Ramsey quantifiers can be used to encode liveness properties, assume
that $R \subseteq \R^k \times \R^k$ is the reachability relation
(i.e.\ transitive closure) of some transition relation $T \subseteq \R^k \times
\R^k$. Then $T$ terminates iff both
$\exists \vec z \colon Start(\vec z) \wedge 
\exists^{\text{ram}} \vec x,\vec y\colon R(\vec z,\vec x) \wedge 
R(\vec x,\vec y)$ and
$\exists \vec x,\vec y \colon Start(\vec x) \wedge R(\vec x,\vec y)
\wedge R(\vec y,\vec y)$
are not satisfied.

It was shown in \cite{ramseyquantifierslineararithmetics} that Ramsey
quantifiers can be eliminated in polynomial time for the existential fragments 
of LIA (Linear Integer Arithmetic), LRA (Linear
Real Arithmetic) and LIRA (Linear Integer Real Arithmetic). Despite this, only a
very prototypical implementation of the algorithms
\cite{ramseyquantifierslineararithmetics} exists for LIA and LRA, and 
no complete implementation so far exists for LIRA. In particular, scalability test was
done only on a small micro-benchmark. Furthermore, the current implementation 
depends very much on the Z3 Python API, and the user cannot
simply specify the constraint in an SMT-LIB-like format, which can be reduced to
any SMT-solver.

\begin{figure}[t]
	\centering
	\includegraphics[width=0.5\textwidth]{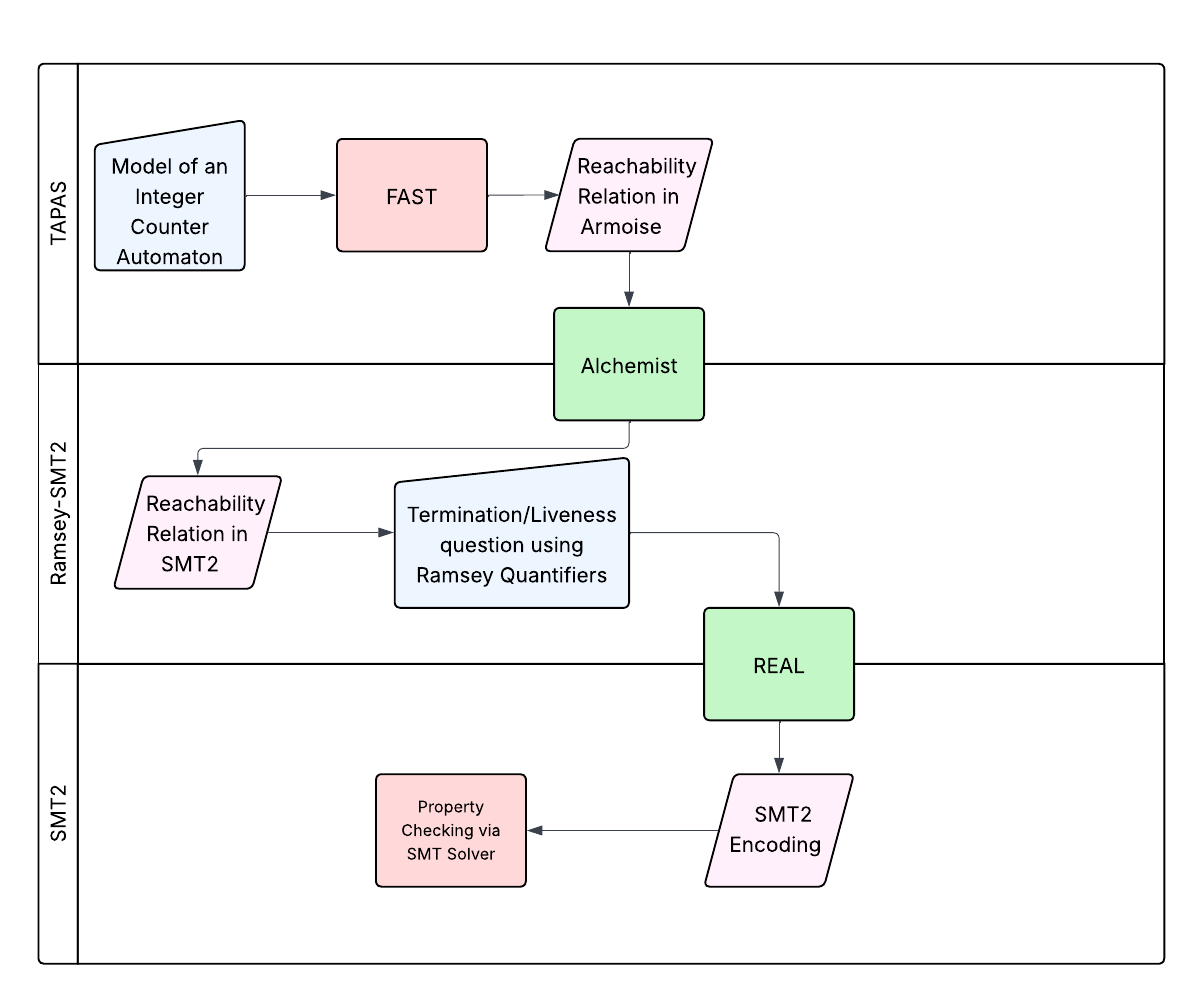}
	\caption{Toolchain}
	\label{fig:toolchain}
\end{figure}

\paragraph{Contributions.} In this paper, we present REAL --- Ramsey Elimination for Arithmetic Logic --- a solver-agnostic elimination procedure for Ramsey quantifiers. REAL allows constraints to be specified in an extension of SMT-LIB with Ramsey quantifiers and significantly improves elimination speeds. REAL also improves on resulting formula sizes and solving times (see \zcref{fig:elim_comp}). Finally, REAL outputs the resulting formula as an SMT-LIB file, which can then be solved by any SMT-solver supporting LIA, LRA, and LIRA.


To demonstrate efficacy of the tool, we apply REAL to liveness verification.
More precisely, we use FAST(er) \cite{FASTer} --- which uses the Talence 
Presburger Arithmetic Suite (TaPAS) \cite{TaPAS} --- to compute the reachability
relations of \emph{integer counter systems} in Armoise, a language for
Presburger-definable sets of integer vectors (equivalently, relations over
numbers). In particular, numerous benchmarks
are available, which encode parameterized distributed protocols. We implemented
\emph{Alchemist}, which transpiles such relations into SMT-LIB \cite{SMTLIB}. 
Combined with REAL, we successfully verify liveness over parameterized 
distributed protocols (using the toolchain described in \zcref{fig:toolchain}),
for which FAST(er) successfully computes reachability relations, including
McCarthy91. See \zcref{tab:liveness_verification}.

%

\paragraph{Illustrating example.}

We illustrate on a simple example how REAL can be used to prove termination.
More involved examples can be found in \zcref{sec:experiments} and \cite{ramseyquantifierslineararithmetics}.
The following program has two integer variables $x_1, x_2$ and
as long as both have values greater 0, the program nondeterministically decrements one of them.

\begin{algorithm}[H]
	\Int $x_1 \gets$ input-int()\;
	\Int $x_2 \gets$ input-int()\;
	\While{$x_1 > 0 \wedge x_2 > 0$}{
		\Either
		$x_1 \gets x_1 - 1$
		\Or
		$x_2 \gets x_2- 1$ 
	}
	\caption{Example of a terminating program}
	\label{fig:program}
\end{algorithm}

We observe that an overapproximation of the reachability relation after at least one iteration of the while loop can be expressed by the LIA formula
\[\varphi := [(y_1 < x_1 \wedge y_2 \le x_2) \vee (y_1 \le x_1 \wedge y_2 < x_2)] \wedge y_1 \ge 0 \wedge y_2 \ge 0\]
where $y_1,y_2$ are the new values of $x_1,x_2$.
Now, REAL in combination with an SMT-solver can be used to verify the absence of both an infinite clique and a loop in $\varphi$,
which proves termination of the program.

	\section{Implementation}\label{sec:implementation}
	\subsection{Alchemist}
	\href{https://github.com/DarkVanityOfLight/alchemist}{Alchemist}
	is a transpiler for the output of FASTer, which is a subset of Armoise, to SMT-LIB.
	We refer to \zcref{app:armoise_example} for an Armoise example and its transpiled version.
	
	Armoise is designed to describe semilinear sets and common operations on them,
	such as union, intersection, and complement.
	It also supports set comprehensions with arbitrary Presburger guards.
	
	The translation from Armoise to SMT-LIB reduces to the standard translation from semilinear sets to Presburger arithmetic 
	\cite{GinsburgSeymour1966SPfa}, together with the elimination
	of any existential quantifiers introduced in the process, 
	which requires inlining all identifier definitions.
	

	
	\subsection{REAL}
	The algorithm presented in \cite{ramseyquantifierslineararithmetics} takes a formula $\psi$ as in \zcref{eq:ramsey},
	where $\varphi$ is an existential formula in LIRA, and 
	computes a formula equivalent to $\psi$ that only contains existential quantifiers, i.e.,
	where the Ramsey quantifier is eliminated.
	To this end, the existentially quantified variables in $\varphi$ are first replaced with additional Ramsey quantified variables.
	Then \cite{ramseyquantifierslineararithmetics} identifies sufficient and necessary conditions
	that can be formulated in existential LIRA for the existence of an infinite clique.
	For example, let
	\[\varphi(\vec x,\vec y,z) := y_1 > x_1 \wedge x_1 + x_2 < z\]
	be a formula in LIA. 
	One can observe that $\exists^{\text{ram}} \vec x, \vec y \colon \varphi(\vec x,\vec y,z)$ is equivalent to
	the existence of a sequence $\vec a , \vec a + \vec b, \vec a + 2\vec b,\dots$ such that
	$a_1+a_2 < z$, $b_1 > 0 $, and $b_1 + b_2 \le 0$, 
	which can be stated by the existential LIA formula
	\[\exists \vec x,\vec y \colon x_1+x_2 < z \wedge y_1 > 0 \wedge y_1 + y_2 \le 0.\]
	Clearly, any such sequence $\vec a , \vec a + \vec b, \vec a + 2\vec b,\dots$ is also an infinite clique of $\varphi$.
	For the converse, let $\vec a_1, \vec a_2, \dots$ be an infinite clique of $\varphi$ for the valuation $z \mapsto c$.
	Let $\vec a_i = (a_{i,1},a_{i,2})$.
	Since $a_{i,1} + a_{i,2} < c$ and $a_{i,1},a_{i,2}$ are integers for all $i \ge 1$, 
	there is $k \ge 1$ such that $a_{k,1} + a_{k,2}  \ge a_{j,1} + a_{j,2} $ for all $j \ge 1$.
	Thus, if we define $\vec a := \vec a_k$ and $\vec b :=  \vec a_{k+1} - \vec a_k$,
	we have that $a_1+a_2 < c$, $b_1 > 0 $, and $b_1 + b_2 \le 0$ as required.
	
	\href{https://github.com/DarkVanityOfLight/REAL}{REAL} improves the prototype implementation in three main categories:
	\begin{enumerate}[(i)]
		\item Enabling input formulas from a convenient extension of SMT-LIB (removing previous strict restrictions) 
		and producing formulas in SMT-LIB format,
		\label{itm:SMT-LIB}
		\item Using more efficient data structures and procedures to speed up the elimination process 
		(e.g.\ caching and sharing (sub)formulas, more efficient atom shape normalization, combining duplicate tree walks),
		\item Producing more compact and structured formulas, e.g., by handling integer equalities and non-strict real inequalities directly, sharing common subexpressions in the Boolean abstraction, and
		replacing unnecessary theory variables with Boolean variables where possible.
	\end{enumerate}
	
	%
	%
	
	For \ref{itm:SMT-LIB}, REAL extends PySMT’s \cite{pysmt2015} abstract syntax tree with the Ramsey quantifier, enabling both parsing from and exporting to an \emph{extended SMT-LIB syntax}.  
	We extend the SMT-LIB grammar (following the conventions in \cite{SMTLIB}) as follows, where \texttt{...} denotes existing production rules:
	\begin{align*}
		\textcolor{green}{\langle \text{term} \rangle} &::= \ldots \\
		&\mid \textcolor{blue}{(\textbf{ramsey}\ (}
		\textcolor{green}{\langle \text{sorted\_var} \rangle}^+
		\textcolor{blue}{)}\  \textcolor{blue}{(} 
		\textcolor{green}{\langle \text{sorted\_var} \rangle}^+ 
		\textcolor{blue}{)}\  
		\textcolor{green}{\langle \text{term} \rangle} 
		\textcolor{blue}{)} \\
		&\mid \textcolor{blue}{(\textbf{ramsey}}\ 
		\textcolor{green}{\langle \text{sort} \rangle}\ 
		\textcolor{blue}{(} \textcolor{green}{\langle \text{symbol} \rangle}^+ \textcolor{blue}{)}\ 
		\textcolor{blue}{(} \textcolor{green}{\langle \text{symbol} \rangle}^+ \textcolor{blue}{)}\ 
		\textcolor{green}{\langle \text{term} \rangle} 
		\textcolor{blue}{)}
	\end{align*}
	Here, blue marks terminal symbols and green marks non-terminals. The first rule is for mixed quantifiers, and the second offers a shorthand for integer or real-only quantifiers.
	The new symbol \textcolor{blue}{\textbf{ramsey}} has the semantics of $\exists^{\text{ram}}$ as defined above.
	

	\section{Benchmarks and Experiments}\label{sec:experiments}
	Experiments were performed on an Arch Linux machine (AMD Ryzen 7 2700X, 32 GiB RAM).
	Satisfiability was checked using Z3 v4.15.1,
	while FASTer utilized the PresTAF solver.
	Tool timings were recorded via the time command.
	While REAL and Alchemist use the inbuilt timing reports.
	
	\subsection{Case Studies}
	
	We demonstrate our approach using the McCarthy 91 function \cite{mccarthy},
	a classic verification benchmark defined as:
	$$ M(n) = 
	\begin{cases}
		n - 10 & \text{if } n > 100 \\
		M(M(n + 11)) &\text{if } n \leq 100
	\end{cases}
	$$
	To verify the absence of infinite cliques for $n \ge 0$,
	we model the function in FASTer using three counters: $n$ (value), $c$ (recursion depth),
	and $s$ (state: $q_0$ evaluating, $q_1$ applying, $q_2$ halting).
	
	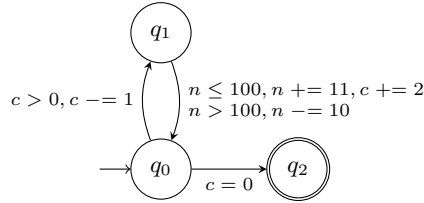
\begin{figure}
		\centering
		\usetikzlibrary{automata,positioning}
		\begin{tikzpicture}[
			node distance=1cm,          
			every state/.style={
				minimum size=0.8cm,      
				inner sep=2pt,           
				font=\small              
			},
			auto,
			every node/.style={font=\scriptsize}, 
			initial text=                
			]
			\node[state, initial] (q0) {$q_0$};
			\node[state] (q1) [above =of q0] {$q_1$};
			\node[state, accepting] (q2) [right=of q0] {$q_2$};
			
			\path[->, >=stealth]
			(q0) edge[bend left=20]  node {$c>0, c\mathrel{-}=1$} (q1)
			(q0) edge[below]         node {$c=0$}         (q2)
			(q1) edge[bend left=20, align=left] node {
				$n\le100, n\mathrel{+}=11, c\mathrel{+}=2$\\
				$n>100, n\mathrel{-}=10$
			} (q0);
		\end{tikzpicture}
		\caption{Transitions for McCarthy91}
		\label{fig:transitions}
	\end{figure}
	
	Adding the transitions of the McCarthy function results in the
	automaton as drawn in \zcref{fig:transitions}.
	
	To produce a reachability relation $R$ rather than a set using FASTer, we double the variables (adding primed versions).
	Transitions update the primed variables while original variables remain constant.
	Initializing the search where each variable equals its primed version allows FASTer to compute the transition relation across arbitrary steps.
	After translating using Alchemist it remains to show how to model the desired property using the Ramsey Quantifier.
	Let $R(n, c, s, n', s', p')$ be the relation generated. 
	
	Even though we could verify the absence of an infinite clique
	for the entire state space, we will for demonstrative purposes show how to specify a valid starting region:
	
	$$ \exists n_f : \exists^{ram} (n, c, s), (n', c', s'): n_f \ge 0 \land R(n_f, 0, 1, n, c, s) \land R(n, c, s, n', c', s')$$
	
	The solver returns UNSAT after elimination, verifying the absence of an infinite clique in $M(n)$ for any given $n\ge 0$
	
	Other benchmarks include:
\begin{description}
	\item[Sliding Window Protocol:] We verify liveness of a TCP-like protocol (with window size 3) to ensure continuous progress.
	\item[Lamport’s Bakery Algorithm \cite{bakery}:]
	A two-process mutual exclusion model.
	Verifying absence of an infinite clique starving one process fails without fairness assumptions.
	\item[Cache Coherence:] 
	We verify the convergence to a coherent state for five protocols, including SYNAPSE, Berkeley, and Dragon \cite{SYNAPSE/Berkley/DRAGON}.
	As well as versions of futurbus and MOESI.
\end{description}

	\begin{figure}[t]
		\centering
		\begin{minipage}{0.48\textwidth}
			\centering
			\resizebox{\textwidth}{!}{%
				\begin{tabular}{llllllr}
					\toprule
					Benchmark       & FAST     & Alchemist & REAL    & Solving  & SAT \\
					\midrule
					sliding\_window & 421.67ms &  0.164s   &  0.076s & 82.61ms  & True \\
					lamport\_bakery & 1.83s    &  9.505s   & 18.629s & 6.23s    & True \\
					mccarthy91      & 145.29ms &  2.282s   &  7.834s & 853.53ms & False \\
					Berkley         & 191.99ms &  0.002s   &  0.003s & 13.88ms  & False \\
					DRAGON          & 2.07s    &  0.457s   &  0.276s & 23.71ms  & False \\
					futurbus        & 125.95s  & 16.972s   & 28.570s & 1.46s    & False \\
					MOESI           & 409.04ms &  0.103s   &  0.052s & 18.04ms  & False \\
					SYNAPSE         & 78.53ms  &  0.017s   &  0.016s & 13.05ms  & False \\
					\bottomrule
				\end{tabular}
			}
		\end{minipage}
		\hfill
		\begin{minipage}{0.48\textwidth}
			\centering
			\resizebox{\textwidth}{!}{%
				\begin{tabular}{llrrr}
					\toprule
					Benchmark & \makecell{Armoise \\ Formula} & \makecell{SMT \\ Pre-elimination} & \makecell{SMT \\ Post-elimination} \\
					\midrule
					sliding\_window & 5795  & 82254    & 18825 \\
					lamport\_bakery & 54348 & 25120657 & 5150912 \\
					mccarthy91      & 3385  & 3984717  & 1701846 \\
					Berkley         & 98    & 35       & 397     \\
					DRAGON          & 20318 & 186122   & 44126   \\
					futurbus        & 83510 & 27605000 & 5433832 \\
					MOESI           & 5941  & 21606    & 9125    \\
					SYNAPSE         & 946   & 2285     & 2709    \\
					\bottomrule
				\end{tabular}
			}
		\end{minipage}
		\caption{Liveness verification results with execution times (left) and formula sizes measured in tree nodes (right)}
		\label{tab:liveness_verification}
	\end{figure}

	\zcref{tab:liveness_verification} shows our pipeline efficiently handles liveness questions, producing concise SMT encodings and enabling practical verification. The original prototype lacks SMT-LIB support, so direct comparison with REAL is not possible, we compare them on different benchmarks below.
	
	\subsection{Elimination}
	We evaluate the performance of REAL on the following parameterized benchmarks.
	Let $\vec{x}, \vec{y} \in \mathbb{K}^{D}$ with $\mathbb{K} \in \{\mathbb{Z},\mathbb{R}\}$.
	
	\[\begin{array}{ll}
		\text{Sorted Chain:} & \exists^{\text{ram}} \vec{x}, \vec{y} : \bigwedge_{i=0}^{D-1} (y_i > x_i) \wedge \bigwedge_{i=0}^{D-2} (y_i < y_{i+1}) \\
		\text{Average:} & \exists^{\text{ram}} \vec{x}, \vec{y} : \bigwedge_{i=0}^{D-1} \Big(D \cdot y_i > \sum_{j=0}^{D-1} x_j \Big) \\
		\text{Cyclic Dependency:} & \exists^{\text{ram}} \vec{x}, \vec{y} : (y_0 > y_{D-1}) \wedge \bigwedge_{i=1}^{D-1} (y_i > x_i + y_{i-1}) \\
		\text{Bounding Box:} & \exists \vec{z}_1, \vec{z}_2:\exists^{\text{ram}} \vec{x}, \vec{y} : \bigwedge_{i=0}^{D-1} (x_i < z_{1,i} < y_i < z_{2,i})
	\end{array}\]
	
	\begin{figure}[t]
		\centering
		\begin{subfigure}[b]{0.48\textwidth}
			\centering
			\includegraphics[width=\textwidth]{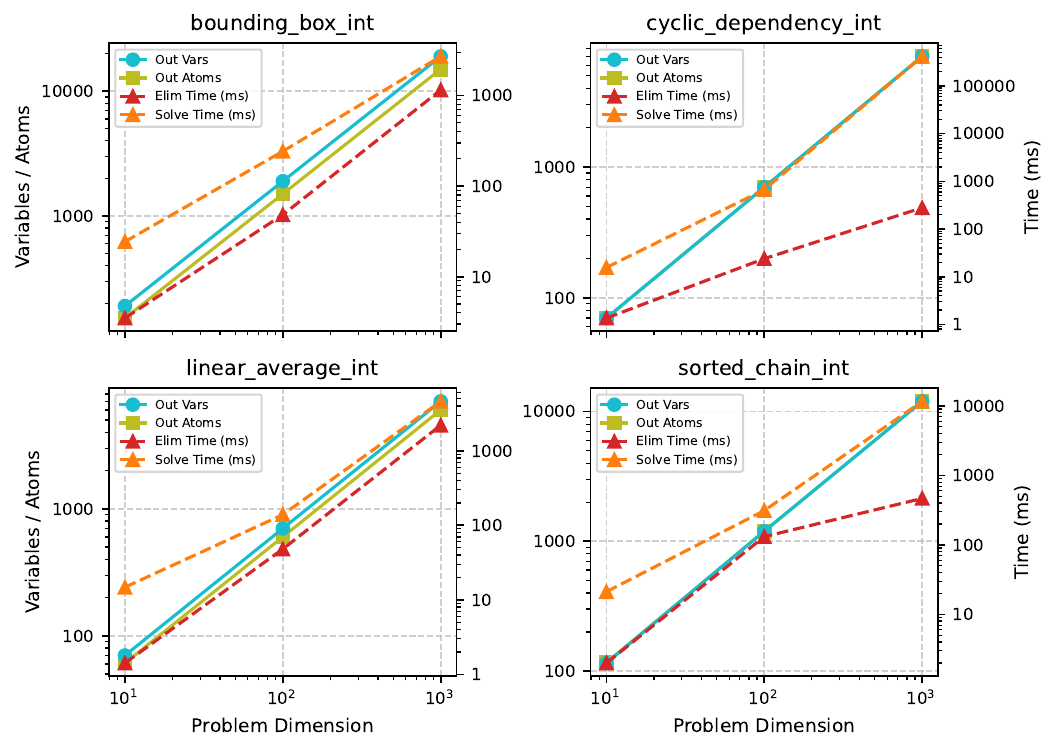}
			\caption{Elimination over the integers}
			\label{fig:int_elim}
		\end{subfigure}
		\hfill
		\begin{subfigure}[b]{0.48\textwidth}
			\centering
			\includegraphics[width=\textwidth]{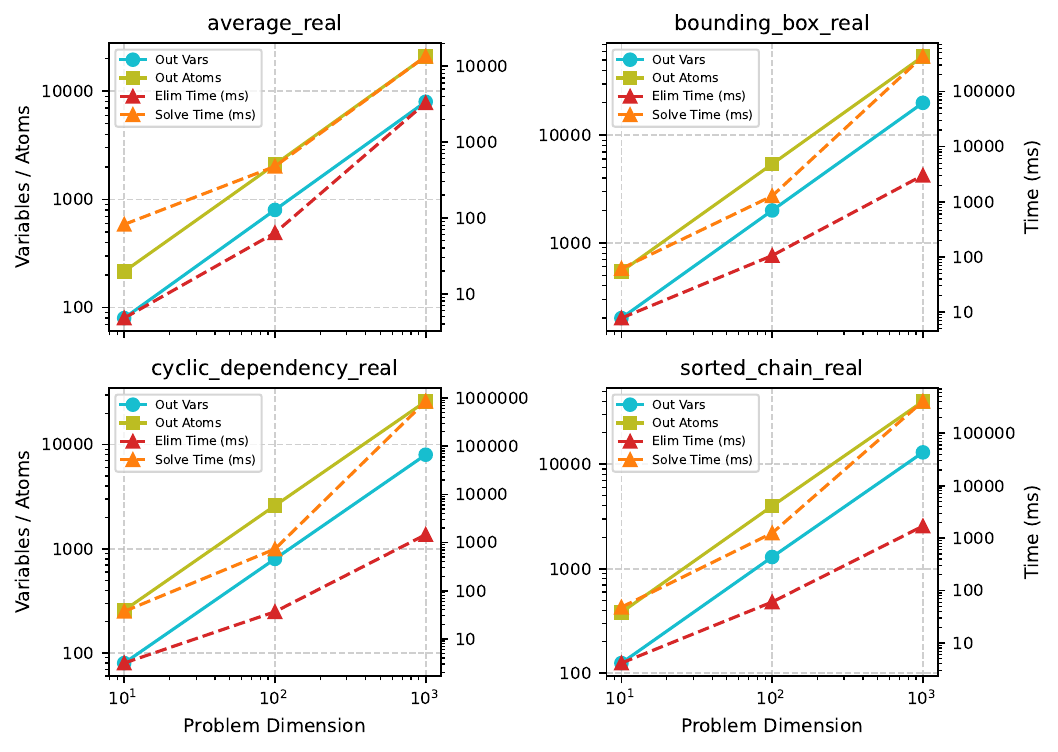}
			\caption{Elimination over the reals}
			\label{fig:real_elim}
		\end{subfigure}
		\caption{Experimental results of REAL on parameterized benchmarks (see \zcref{app:elim_data})}
		\label{fig:elim}
	\end{figure}
	
	The results in \zcref{fig:int_elim,fig:real_elim} highlight two key trends. First, elimination scales almost linearly with the problem dimension across all benchmark families and is consistently dominated by solving times. Even formulas with thousands of atoms and variables complete in the sub-second to low-second range.

	\subsection{Comparison}
	To evaluate performance improvements over the original prototype (prot), we ran additional benchmarks using a modified version of the Average benchmark for $D=100$.
	
	\begin{itemize} 
		\item $\text{Real case: } \exists^{\text{ram}} \vec{x}, \vec{y} : \bigwedge_{i=0}^{D-1} \Big( D \cdot y_i \ge \sum_{j=0}^{D-1} x_j \Big)$
		\item $\text{Integer case: } \exists^{\text{ram}} \vec{x}, \vec{y} : \bigwedge_{i=0}^{D-1} \Big( D \cdot y_i = \sum_{j=0}^{D-1} x_j \Big)$
	\end{itemize}
	
	\begin{figure}[t]
		\centering
		\begin{subfigure}[b]{0.48\textwidth}
			\centering
			\includegraphics[width=\textwidth]{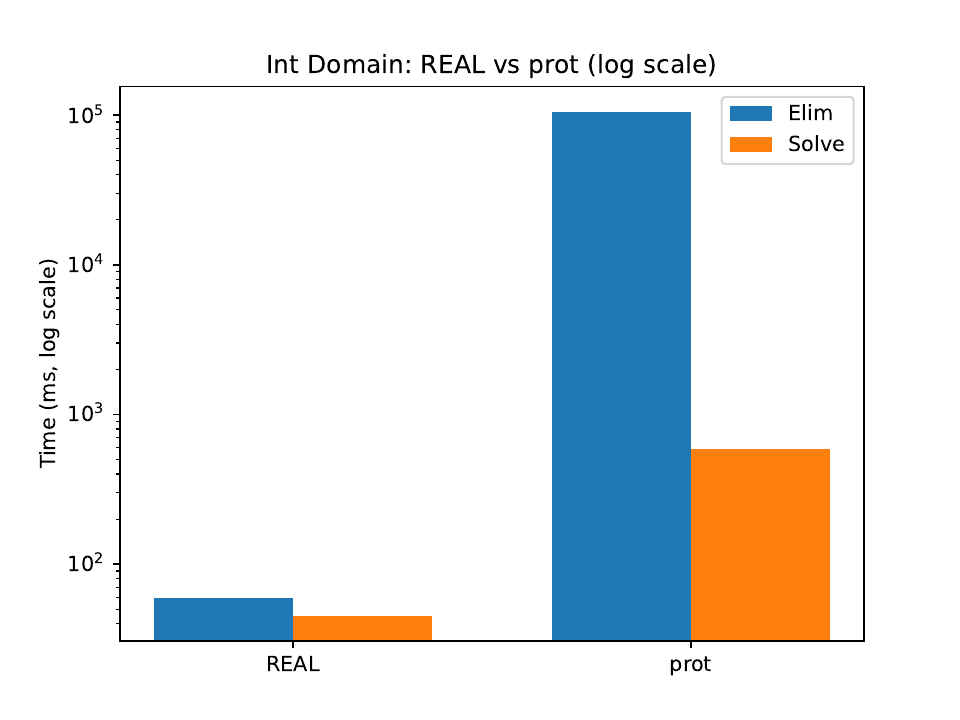}
			\caption{Elimination over the integers}
			\label{fig:int_elim_comp}
		\end{subfigure}
		\hfill
		\begin{subfigure}[b]{0.48\textwidth}
			\centering
			\includegraphics[width=\textwidth]{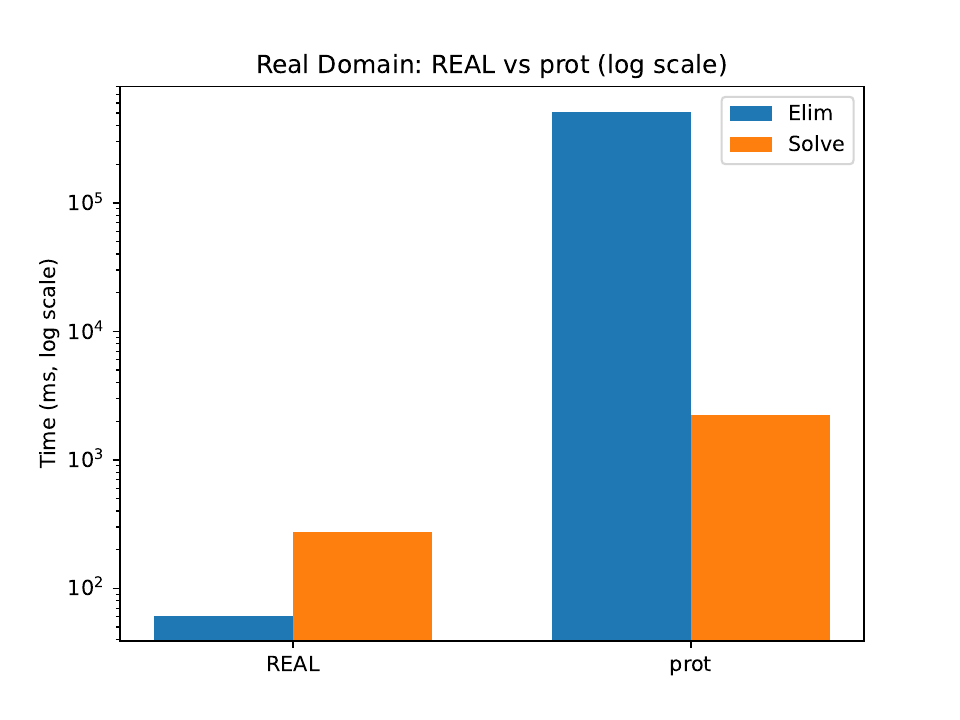}
			\caption{Elimination over the reals}
			\label{fig:real_elim_comp}
		\end{subfigure}
		\caption{Comparison of REAL and the prototype in terms of elimination and solving times (see \zcref{app:comp_data})}
		\label{fig:elim_comp}
	\end{figure}
	
	The results in \zcref{fig:elim_comp} show that the new elimination method outperforms the previous implementation in both elimination and solving times by several magnitudes. In the integer case, the generated output is so simple and easy to prove unsatisfiable that the solver is faster than our elimination procedure.
	However, as demonstrated above, the typical trend is that elimination time is dominated by solving time. 
	
	\section{Conclusion and Future Work}
	We presented a modular verification pipeline that combines Armoise specifications, the Alchemist transpiler, and REAL. Benchmarks on classical and practical examples demonstrate that the approach efficiently handles infinite-state systems.
	
	The case of mixed linear integer real arithmetic still leaves room for improvement, 
	as the overhead compared to pure LIA and LRA remains substantial, 
	due to the separation of integer and real variables in the current approach.
	Moreover, replacing FASTer with a tool that directly targets reachability relations, or streamlining FASTer’s ergonomics, could significantly reduce the size and redundancy of SMT encodings.
	Finally, we propose to extend the pipeline to counter systems over the reals (and the mixed case).
	
	\begin{credits}
		\subsubsection{\ackname} 
		We thank anonymous reviewers for their helpful feedback.
		
		Funded by the Deutsche Forschungsgemeinschaft (DFG, German Research Foundation) – \href{https://gepris.dfg.de/gepris/projekt/522843867}{522843867}.
		
		\subsubsection{Data-Availability Statement.}
		The experimental results of this paper may be reproduced using the artifact on Zenodo \cite{artifact}.
		For reuse of our tools, we refer to the GitHub repositories of
		\href{https://github.com/DarkVanityOfLight/alchemist}{Alchemist} and \href{https://github.com/DarkVanityOfLight/REAL}{REAL}.
	\end{credits}

	%
	%
	%
	\bibliographystyle{splncs04}
	\bibliography{references.bib}

	\appendix
	\renewcommand{\theHsection}{A\arabic{section}} 
	
	\section{Elimination Data}\label{app:elim_data}
	\small
	\setlength{\tabcolsep}{4pt}
	\renewcommand{\arraystretch}{0.9}
	\noindent\textbf{bounding\_box\_int}\\
	\begin{tabular}{rrrrrrr}
		\toprule
		Dimension & In Vars & Out Vars & In Atoms & Out Atoms & Elim (ms) & Solve (ms) \\
		\midrule
		10 & 40 & 190 & 30 & 152 & 3.49 & 24.35 \\
		100 & 400 & 1900 & 300 & 1502 & 48.25 & 242.07 \\
		1000 & 4000 & 19000 & 3000 & 15002 & 1150.99 & 2703.15 \\
		\bottomrule
	\end{tabular}
	
	\vspace{1em}
	
	\noindent\textbf{cyclic\_dependency\_int}\\
	\begin{tabular}{rrrrrrr}
		\toprule
		Dimension & In Vars & Out Vars & In Atoms & Out Atoms & Elim (ms) & Solve (ms) \\
		\midrule
		10 & 19 & 70 & 10 & 70 & 1.36 & 15.62 \\
		100 & 199 & 700 & 100 & 700 & 23.95 & 670.54 \\
		1000 & 1999 & 7000 & 1000 & 7000 & 279.75 & 415447.81 \\
		\bottomrule
	\end{tabular}
	
	\vspace{1em}
	
	\noindent\textbf{linear\_average\_int}\\
	\begin{tabular}{rrrrrrr}
		\toprule
		Dimension & In Vars & Out Vars & In Atoms & Out Atoms & Elim (ms) & Solve (ms) \\
		\midrule
		10 & 20 & 70 & 10 & 61 & 1.42 & 14.77 \\
		100 & 200 & 700 & 100 & 601 & 48.12 & 139.05 \\
		1000 & 2000 & 7000 & 1000 & 6001 & 2200.84 & 4596.59 \\
		\bottomrule
	\end{tabular}
	
	\vspace{1em}
	
	\noindent\textbf{sorted\_chain\_int}\\
	\begin{tabular}{rrrrrrr}
		\toprule
		Dimension & In Vars & Out Vars & In Atoms & Out Atoms & Elim (ms) & Solve (ms) \\
		\midrule
		10 & 20 & 115 & 19 & 116 & 1.99 & 21.14 \\
		100 & 200 & 1195 & 199 & 1196 & 131.37 & 312.04 \\
		1000 & 2000 & 11995 & 1999 & 11996 & 466.17 & 11614.51 \\
		\bottomrule
	\end{tabular}

	\noindent\textbf{average\_real}\\
	\begin{tabular}{rrrrrrr}
		\toprule
		Dimension & In Vars & Out Vars & In Atoms & Out Atoms & Elim (ms) & Solve (ms) \\
		\midrule
		10 & 20 & 80 & 10 & 216 & 4.77 & 81.98 \\
		100 & 200 & 800 & 100 & 2106 & 63.41 & 477.24 \\
		1000 & 2000 & 8000 & 1000 & 21006 & 3280.07 & 13466.06 \\
		\bottomrule
	\end{tabular}
	
	\vspace{1em}
	
	\noindent\textbf{bounding\_box\_real}\\
	\begin{tabular}{rrrrrrr}
		\toprule
		Dimension & In Vars & Out Vars & In Atoms & Out Atoms & Elim (ms) & Solve (ms) \\
		\midrule
		10 & 40 & 200 & 30 & 542 & 7.81 & 61.66 \\
		100 & 400 & 2000 & 300 & 5402 & 106.06 & 1276.86 \\
		1000 & 4000 & 20000 & 3000 & 54002 & 3022.24 & 431418.32 \\
		\bottomrule
	\end{tabular}
	
	\vspace{1em}
	
	\noindent\textbf{cyclic\_dependency\_real}\\
	\begin{tabular}{rrrrrrr}
		\toprule
		Dimension & In Vars & Out Vars & In Atoms & Out Atoms & Elim (ms) & Solve (ms) \\
		\midrule
		10 & 19 & 80 & 10 & 257 & 3.13 & 37.06 \\
		100 & 199 & 800 & 100 & 2597 & 36.45 & 733.26 \\
		1000 & 1999 & 8000 & 1000 & 25997 & 1463.06 & 855566.40 \\
		\bottomrule
	\end{tabular}
	
	\vspace{1em}
	\noindent\textbf{sorted\_chain\_real}\\
	\begin{tabular}{rrrrrrr}
		\toprule
		Dimension & In Vars & Out Vars & In Atoms & Out Atoms & Elim (ms) & Solve (ms) \\
		\midrule
		10 & 20 & 125 & 19 & 382 & 4.14 & 48.14 \\
		100 & 200 & 1295 & 199 & 3982 & 60.56 & 1250.29 \\
		1000 & 2000 & 12995 & 1999 & 39982 & 1696.88 & 405891.43 \\
		\bottomrule
	\end{tabular}

	\section{Comparison Data}\label{app:comp_data}
	\begin{table}[H]
		\centering
		\caption{Benchmark Results}
		\label{tab:benchmark}
		\resizebox{\textwidth}{!}{%
			\begin{tabular}{@{}lccccc@{}}
				\toprule
				\textbf{Benchmark} & \textbf{Tool} & \textbf{Elim (ms)} & \textbf{Solve (ms)} & \textbf{Total (ms)} & \textbf{Result} \\ \midrule
				\texttt{linear\_average\_eq\_int} & REAL & 59.71 & 45.37 & 105.08 & Unsat \\
				& Prototype & 105530.52 & 586.09 & 106116.61 & Unsat \\ \midrule
				\texttt{average\_eq\_real} & REAL & 61.35 & 273.79 & 335.14 & Sat \\
				& Prototype & 513192.25 & 2221.74 & 515414.00 & Sat \\ \bottomrule
			\end{tabular}%
		}
	\end{table}

	\begin{table}[H]
		\centering
		\caption{Benchmark Results: Variables and Atoms}
		\label{tab:benchmark_vars_atoms}
		\resizebox{\textwidth}{!}{%
			\begin{tabular}{@{}lccccc@{}}
				\toprule
				\textbf{Benchmark} & \textbf{Tool} & \textbf{In Vars} & \textbf{In Atoms} & \textbf{Out Vars} & \textbf{Out Atoms} \\ \midrule
				\texttt{linear\_average\_eq\_int} & REAL & 200 & 100 & 300 & 301 \\
				& Prototype & 200 & 100 & 2200 & 13000 \\ \midrule
				\texttt{average\_eq\_real} & REAL & 200 & 100 & 800 & 2006 \\
				& Prototype & 200 & 100 & 1900 & 20900 \\ \bottomrule
			\end{tabular}%
		}
	\end{table}
	
	
	\section{Armoise example}\label{app:armoise_example}
	\lstset{
		language=Scala,
		basicstyle=\ttfamily\small,
		keywordstyle=\bfseries,
		commentstyle=\itshape\color{gray},
		columns=flexible,
		keepspaces=true,
		showstringspaces=false,
		numbersep=5pt,
		numbers=left,
		frame=single,
		aboveskip=1em,
		belowskip=1em,
		tabsize=2,
	}
	Definition of a ternary relation in Armoise:
	\begin{lstlisting}
let
	int3 := (int, int, int);
	base := nat * (1, 0, 0) + nat * (0, 1, 0) + nat * (0, 0, 1);
	// Nested let blocks
	evensLarger20oddsSmaller20 := let
		// Arithmetic operations on sets 
		evens := 2*int3;
		odds := (1, 1, 1) + evens;
		// Set comprehensions
		sumSmaller20 := {(x_1, x_2, x_3) in base | x_1 + x_2 + x_3 < 20 };
		// Set operations(difference, intersection, union)
		evensLarger20 := evens \ sumSmaller20;
		oddsSmaller20 := odds && sumSmaller20;
	in
	evensLarger20 || oddsSmaller20;
in
{ (x, y, z) in int3 | (x, y, z) in evensLarger20oddsSmaller20 };
	\end{lstlisting}
	
	\noindent Result of the transpilation to SMT-LIB:
	\begin{lstlisting}
(define-fun R ((x Int) (y Int) (z Int)) Bool
  (and true
  	(or
    	(and
      	(and (= (mod x 2) 0)
            	(= (mod y 2) 0)
            	(= (mod z 2) 0)
            	true)
      	(not
        	(and
          	(and (and (>= x 0)
                    	(>= y 0)
                    	(>= z 0))
                	(< (+ (* 1 (- (* 1 x) 0))
                      	(* 1 (- (* 1 y) 0))
                      	(* 1 (- (* 1 z) 0)))
                  	(* 1 20))))))
    	(and
      	(and (= (mod x 2) 1)
            	(= (mod y 2) 1)
            	(= (mod z 2) 1)
            	true)
      	(and
        	(and (and (>= x 0)
                  	(>= y 0)
                  	(>= z 0))
              	(< (+ (* 1 (- (* 1 x) 0))
                    	(* 1 (- (* 1 y) 0))
                    	(* 1 (- (* 1 z) 0)))
                	(* 1 20))))))))
	\end{lstlisting}
	
\end{document}